\documentclass[authoryear,preprint,review,12pt]{elsarticle}

\usepackage{amssymb}
\usepackage{graphicx}
\usepackage{color}
\newcommand{\Rsun}{\ensuremath{\,{\rm R}_\odot}}                        % Solar radius symbol
\newcommand{\kms}{\,km\,s$^{-1}$}                                       % km/s symbol

\journal{New Astronomy}

\begin{document}
\begin{frontmatter}



\title{The early B-type eclipsing binary GT\,Cephei: a massive triple system?}

\author[l1]{\"{O}. \c{C}ak{\i}rl{\i} \corref{cer}}

\address[l1]{Ege University, Science Faculty, Astronomy and Space Sciences Dept., 35100 Bornova, \.{I}zmir, Turkey \corref{cer}}
\cortext[cer]{Corresponding author. Tel.: +902323111740; Fax: +902323731403 \\
E-mail address: omur.cakirli@gmail.com}

\begin{abstract}
GT\,Cep is a semi-detached close binary system with an orbital period of 4.91 
days, containing a massive star. I have obtained spectroscopic observations and derived 
radial velocities of both components. Combining the analyses of radial velocities and 
available photometric observations we have measured the absolute parameters of both 
components of GT\,Cep. The components are shown to be a B2\,V primary with a mass 
M$_p$=10.70$\pm$0.50 M$_{\odot}$ and radius R$_p$=6.83$\pm$0.19 R$_{\odot}$ and a A0\,IV 
secondary with a mass M$_s$=2.58$\pm$0.14 M$_{\odot}$ and radius R$_s$=7.56$\pm$0.21 R$_{\odot}$. My 
analyses show that GT\,Cep is a classical Algol-type binary with a less massive secondary filling its 
$Roche$ lobe. Using the UBVJHK magnitudes and the interstellar reddening of E(B-V)=0.61 I estimated the 
mean distance to the system as 854$\pm$43\,pc. The O-C residuals have been analyzed as the consequence
of a light-time effect superimposed on an upward parabola. My analysis 
indicates that the eclipsing binary revolves around a third-body with a period 
of about 57.5 yr in an orbit with a radius of 40 $AU$. The lower limit for the 
mass of the third star has been estimated to be 7 M$_{\odot}$ for the 
inclination between $70^{o}$ and $90^{o}$. 
\end{abstract}

\end{frontmatter}

\section{INTRODUCTION }
\label{sec:intro}
One of the most important parameters in stellar astrophysics is 
the mass of stars. Eclipsing binaries with well-defined multi-passband light curves and accurate radial velocities 
for both components provide us with definitive empirical masses, radii, effective temperatures and luminosities. 
In order to better understand the physics of binary systems and test the predictions of theoretical models, it is thus 
important to quantitatively analyse the properties of massive binary systems with 
well-constrained orbital parameters. In this context studies of the rare early B-type 
massive stars has major highlights. 

The relatively bright eclipsing binary GT\,Cep (HD\,217224, HIP113385, V=8.13, B-V=0.34)
was discovered to be an eclipsing binary system by \citet{stro}, who 
derived an orbital period of 4.908756\,d using the 12 times of minima obtained from photographic observations.
They obtained the first photographic light curve of the system and classified it as an Algol-type binary.
A first spectroscopic study was carried out by \citet{pim1964} who obtained the spectroscopic orbit and 
classified the primary component as a B3 star. Later on the photographic study made by 
\citet{1975PZP.....2..171B} who revised the orbital period and obtained a light curve 
containing rather a deep primary minimum and a shallow secondary minimum.  
\citet{1984A&AS...55..403B} obtained UBV light curves and analyzed using the Wood's method. 

Photometric observations of GT\,Cep were also obtained by the Hipparcos satellite 
(GT\,Cep being identified as HIP\,113385 \citep{esa}) and later in the context of the Northern Sky 
Variability Survey \citep{Wozniak04}.
Although GT\,Cep has been studied on many occasions its astrophysical parameters were not firmly established.
On the other hand \citet{2006MNRAS.373..435I} divided the semi-detached Algol-type binaries (SDABs) into two groups. 
The orbital angular momenta of SDABs with periods P$<5$ days and P$>$5 days are 45 and 25 per cent smaller than those detached 
binaries with similar mass. The specific angular momenta of systems with P$>$5 d are larger than than those of P$<5$ d
for the gainers of the same mass. The spins of the mass gaining stars point out a sharp distinction between short and long period orbit
systems at an orbital period of 5 days. The orbital period of GT\,Cep is very close to this discriminating period.   

This paper is organized as follows. I present new spectroscopic observations and radial 
velocities of both components of the eclipsing pair. By analysing the previously published 
light curves and the new radial velocities I obtain orbital parameters for the components. Combining 
the results of these analyses we obtain absolute physical parameters of both components. In 
addition, I conclude with a brief discussion of the system's evolutionary status.

%%%%%%%%%%%%%%%%%%%%%%%%%%%%%%%%%%%%5
\section{OBSERVATIONS}
The present study is the result of a collaboration in which two data sets were obtained 
at two different in the roughly same latitude observatories, using the two telescopes and spectrographs.

The first dataset was obtained  with the REOSC Echelle spectrograph mounted 
on the 182\,cm telescope at the Asiago Observatory in Italy, with exposure time ranging from 
30 to 45 minutes. The instrument covers the spectral domain between 3900 and 7300 \AA, divided 
into 27 orders. The average signal-to-noise ratio (S/N) and resolving power 
$\lambda / \Delta\lambda$ were about 120 and $\sim$ 50\,000, respectively. Four {\it \'{e}chelle} spectra of GT\,Cep were 
taken from March 17, 2009 to July 8, 2011. 

12 {\it \'{e}chelle} spectra of GT\,Cep were collected with the Turkish Faint Object 
Spectrograph Camera (TFOSC)\footnote{http://tug.tug.tubitak.gov.tr/rtt150\_tfosc.php} 
attached to the 1.5 m telescope between August 22, 2011 and August 2, 2013, under good seeing conditions. Further 
details on the telescope and the spectrograph can be found at http://www.tug.tubitak.gov.tr. The 
wavelength coverage of each spectrum was 4000-9000 \AA~in 12 orders, with a resolving power 
of $\lambda$/$\Delta \lambda$ $\sim$7\,000 at 6563 \AA~and an average signal-to-noise ratio 
(S/N) was $\sim$120. I also obtained high S/N spectra of the early type standard stars
1\,Cas (B0.5\,IV), HR\,153 (B2\,IV), $\tau$ Her (B5\,IV), 21\,Peg (B9.5\,V) and $\alpha$Lyr (A0\,V)   
for use as templates in derivation of the radial velocities.

I applied the same reduction procedure to both datasets. The electronic bias was removed 
from each image and I used the 'crreject' option for cosmic ray removal. Thus, the resulting 
spectra were largely cleaned from the cosmic rays. The {\it \'{e}chelle} spectra were extracted and 
wavelength calibrated by using Fe-Ar lamp source with help of the IRAF {\footnote{IRAF is distributed by the $National~Optical~
Astronomy$ Observatory, which is operated by the Association of Universities for Research in Astronomy,Inc. (AURA), under 
cooperative agreement with the National Science Foundation} \sc echelle package} \citep{Sim74}.

%%%%%%%%%%%%%%%%%%%%%%%%%%%%%%%%%%%%%%%%%5

\section{RADIAL VELOCITIES AND ATMOSPHERIC PARAMETERS}
\subsection{Period Determination}
A total of 22 times of mid-primary minimum and one secondary of GT\,Cep were collected from the literature 
and listed in Table\,1. The starting epoch and orbital period are taken from the $Hipparcos$ and \citet{Kre04}, respectively. Therefore 
the following ephemeris was used to determine the cycle number and $O-C$(I) residuals,

\begin{equation}
Min I(HJD)=2\,448\,503.19+4^d.9087946\times E.
\end{equation}

The O-C(I) residuals, indicating the differences between observed times of mid-eclipses and calculated
ones using this ephemeris are listed in the third column of Table\,1. These residuals for all the times of mid-eclipses
of the GT\,Cep are plotted against the epoch numbers in the top panel of Fig.\,1. The trend of the $O-C$(I) residulas can be described by an 
upward parabolic curve superimposed on a sine-like variation. It is obvious that the change of the $O-C$(I) residuals is a result of at least two 
separate causes. Because  GT\,Cep is a semi-detached Algol-type binary, the system could be transferring mass from less massive component 
to the more massive primary leading to an upward parabolic change of orbital period, i.e., indicating that the orbital period is continuously increasing.

 Recently \citet{Lia10} suggested that cyclic period changes are a common phenomenon in close binary systems. Cyclic variations
in the orbital periods are usually explained by magnetic activity in one or both components,  by an apsidal motion, and or by the light-travel time 
effect around common-center with a third-body. Since GT\,Cep is composed of a B2 V and a A0 IV star that contain convective core and radiative 
atmosphere. This suggests that the cyclic changes in the $O-C$ residuals can not be originated from the magnetic activity cycle mechanism.  
We can, therefore, easily ignore the possibility of solar-like activity in both comments. Both the light and radial velocities of the system
point out a circular orbit for the system. In addition, the O-C(I) residual obtained for the mid-secondary
eclipse seem to follow the same trend as the primary minimum. Therefore, we may rule out the apsidal motion
as a possible cause of orbital period change. Therefore such a sinusoidal/cyclic change in the orbital period of  GT\,Cep can only be explained by 
an orbital motion around a third-body. We analyzed the O-C(I) residuals under an assumption of a combination of mass-transfer and third-body, i.e. the 
eclipsing pair is orbiting around a third-body. We may compute the times of light minimum with a formula as 

\begin{equation}
T_{\rm ec}=T_{\rm 1}+ P_{\rm 1}\times E+ Q\times E^2 +\delta T.           
\end{equation}

where  T$_1$ is the starting epoch, E is the integer eclipse number and P$_1$ is the orbital period of the eclipsing pair. While the third-term 
represents the parabolic change in the $O-C$ residuals the time delay or advance of any observed eclipse is caused by the influence 
of a third-body can be represented by a fourth-term. The light-time effect $\delta$T is depended up on the semi-major axis of the eclipsing pair 
around the barycenter, inclination, eccentricity and longitude of the periastron of the third-body orbit. I have used the conventional formulae 
given by \citet{Iba00}. A linear least squares solution was applied to the data and the coefficient of the third-term and the parameters for the 
third-body orbit were obtained. The coefficient of the quadratic-term, originated from the mass transfer between the components is calculated 
as  Q=1.28x10$^{-9}$ $\pm$0.22x10$^{-9}$ days\,cycle$^{-1}$. A secular period increase has been calculated as  dP/dt=1.90x10 $^{-7}$d\,yr$^{-1}$ which corresponds 
to  1.64 sec\,century $^{-1}$, indicating that mass transfer from less massive to the more massive star at a rate of dM/dt=4.4x10$^{-8}$\,M$_{\odot}$\,yr$^{-1}$.

The parameters of the third-body orbit are listed in Table\,2. My calculation suggests that the eclipsing pair revolves around a third-body, in an 
eccentric orbit with e=0.047, and with a period of about 57.5$\pm$ 2.3 years. The projected radius of the orbit of the eclipsing pair around the 
center--of--mass is about 13.80$\pm$0.52 $AU$. Using these values I obtained a mass function as 0.796$\pm$0.060 M$_{\odot}$. The 
O-C(II) residuals were obtained after subtracting the continuous period increase and the light-time effect, and are plotted in the bottom panel of Fig.\,1.

\begin{figure*}
\includegraphics[width=14.5cm,angle=0]{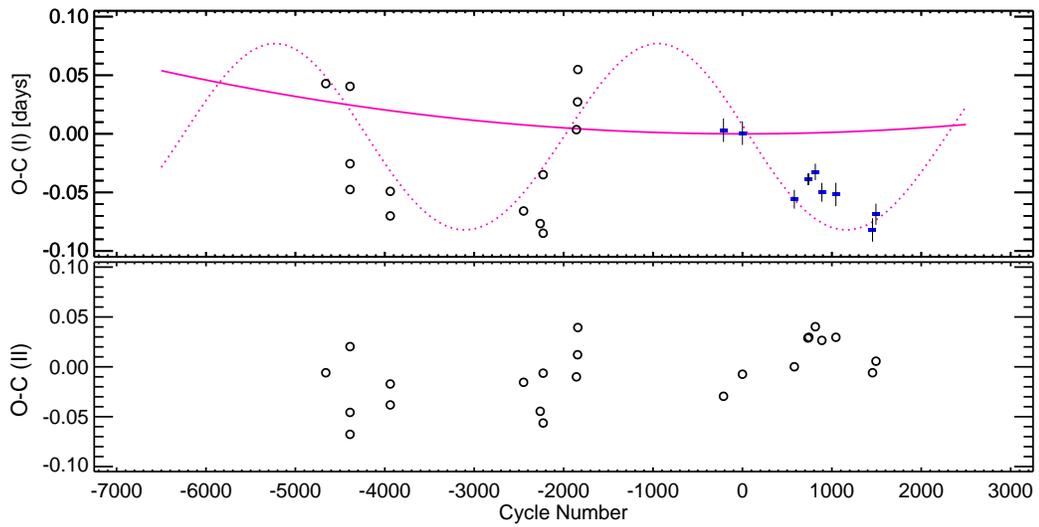}
\caption{The $O-C$(I) diagram for GT\,Cep obtained using the Eq.(1) and its representation 
with a quadratic-term (solid line) and a light-time effect (dotted line). In the bottom panel 
I plotted the  $O-C$(II) residuals, deviations from the upward parabola and a light-time fit, versus 
the epoch numbers. The filled lying-bar with error bars refer to photoelectric times for mid-minima 
and open circles refer to photographic ones.
} \end{figure*}

\begin{table} \begin{center}
\scriptsize
\caption{\label{tab:minima} Literature times of minimum light of GT\,Cep and the observed
minus calculated ($O-C$) values of the data obtained by Eqs.(2) and (3). {\bf References:} (1) (Strohmeier 1962);
(2) (Berthold 1976); (3) (Diethelm 2003); (4) (Kreiner 2004); (5) (Meyer 2006); 
(devil) (VSX-O-C Gateway-http://var.astro.cz/ocgate/ocgate.php); 
 (7) (ESA 1997); (music) (Wozniak et al. 2004). }
\begin{tabular}{lcrrcc} \hline
Time of minimum           &    Cycle    &O-C(I) &O-C(II) &  Type&Reference  \\
(HJD $-$ 2\,400\,000)     &    number   &(days) &(days)  &             \\
\hline
25628.250 & -4660  &  0.043 & -0.006  &pg& 1 \\
26958.443 & -4389  & -0.048 & -0.068  &pg& 1 \\
26958.465 & -4389  & -0.026 & -0.046  &pg& 1 \\
26958.531 & -4389  &  0.040 &  0.020  &pg& 1 \\
29167.378 & -3939  & -0.070 & -0.038  &pg& 1 \\
29167.399 & -3939  & -0.049 & -0.017  &pg& 1 \\
36486.395 & -2448  & -0.066 & -0.016  &pg& 2 \\
37399.420 & -2262  & -0.077 & -0.045  &pg& 2 \\
37561.402 & -2229  & -0.085 & -0.056  &pg& 2 \\
37561.452 & -2229  & -0.035 & -0.006  &pg& 2 \\
39387.562 & -1857  &  0.004 & -0.010  &pg& 6 \\
39451.400 & -1844  &  0.027 &  0.012  &pg& 6 \\
39466.154 & -1841  &  0.055 &  0.039  &pg& 6 \\
47462.5279 & -212   &  0.0024 & -0.0295  &pe& 6 \\
48503.190 &  0     &  0.000 & -0.007  &pe& 7 \\
51350.235 & 580    & -0.056 &  0.000  &pe& 8 \\
52096.389 & 732    & -0.039 &  0.029  &pe& 3 \\
52145.477 & 742    & -0.039 &  0.030  &pe& 3 \\
52503.8253 & 815    & -0.0323 &  0.0403  &pe& 6 \\
52862.150 & 888    & -0.050 &  0.027  &pe& 4 \\
53632.829 & 1045   & -0.051 &  0.030  &pe& 5 \\
55645.4049 & 1455   & -0.0812 & -0.0059  &pe& 6 \\
55834.4069 & 1493.5 & -0.0678 &  0.0057  &pe& 6 \\
\hline \end{tabular} \end{center} \end{table}

\begin{table}
\scriptsize
\centering
\begin{minipage}{85mm}
\caption {Orbital solution for the third component of GT\,Cep.}
\begin{tabular}{@{}lcccccccc@{}c}
\hline
Parameter  & Value   &   $\sigma$	\\						
\hline 
 T$_1$ (HJD)	    		& 2448503.1967  		&0.0011		\\ 
 P$_1$ (day)	    		& 4.908803  		        &0.000003	\\ 
 $A$ (day)	    		& 0.080  		        &0.002		\\
 $e$	    		    	& 0.047  		        &0.005		\\ 
 $\omega$	    		& 41  		                &2		\\  
 $T_3$ (HJD)	    		& 2440956.21  		    	&0.17		\\ 
 $P_3$ (year)	    		& 57.5  		        &2.3		\\   
 $a_{12}\,sin\,i$ (AU)	   	& 13.80          		&0.52		\\   
\hline
\end{tabular}
\end{minipage}
\end{table}

\subsection{Radial velocity}
To derive the radial velocities, the sixteen spectra obtained for the system are cross correlated against the 
template spectra of standard stars 1\,Cas, HR\,153, $\tau$ Her, 21\,Peg and $\alpha$Lyr on an order-by-order
basis using the {\sc fxcor} package in IRAF. The standard stars' spectra were synthetically broadened by 
convolution with the broadening function of \citet{Gra92}. The cross-correlation with the standard star 
$\tau$ Her gave the best result. I have also used $\alpha$Lyr for some spectra.

The spectra showed two distinct cross-correlation peaks in the quadratures, one for each component of the 
binary. Thus, both peaks are fitted independently with a $Gaussian$ profile to measure the velocities and 
their errors for the individual components. If the two peaks appear blended, a double Gaussian was applied 
to the combined profile using {\it de-blend} function in the task. I applied the cross-correlation technique 
to two wavelength regions with well-defined absorption lines of the primary and secondary components. These 
regions (3rd and 4th orders) include the He\,{\sc i} $\lambda$6678 and $\lambda$5876 \AA~ lines, dominant in 
early B-type stars. Here I used as weights the inverse of the variance of the radial velocity measurements 
in each order, as reported by {\sc fxcor}. I have been able to measure radial velocities of both components 
with a precision below than 10\,\kms. 

The heliocentric radial velocities for the primary (V$_p$) and the secondary (V$_s$) components are listed in 
Table\,3, along with the dates of observations and the corresponding orbital phases computed with the new 
ephemeris given in Eq.2. The velocities in this table have been corrected to the heliocentric reference system 
by adopting a radial velocity value for the template stars. The radial velocities are plotted against the 
orbital phase in Fig.\,2 where the empty squares correspond to the primary and the filled squares to the 
secondary star. I have analyzed all the radial velocities for the initial orbital parameters using the {\sc RVSIM} software program 
\citet{kane}. The best fit was obtained for a circular orbit with a mass-ratio of $q=\frac{M_2}{M_1}$=0.2415$\pm$0.0094 and a projected
separation of $asin~i$=28.42 $\pm$0.43 R$_{\odot}$. The orbital solution is presented in Table\,4.
The continuous lines in Fig.\,2 show the computed curves. 

\begin{figure}
\includegraphics[width=12.5cm,angle=0]{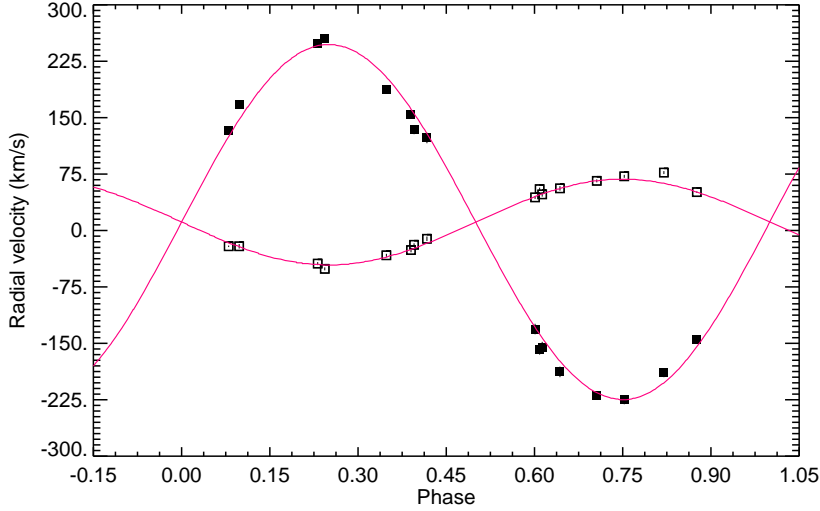}
\caption{Radial velocities for the components of GT\,Cep. Symbols with error bars, generally smaller than the symbol size, show the radial velocity 
measurements for the components of the system (primary: open squares, secondary: filled squares).} \end{figure}

\begin{table}
\scriptsize
\centering
\begin{minipage}{85mm}
\caption{Heliocentric radial velocities of GT\,Cep. The columns give the heliocentric 
Julian date, the orbital phase (according to the ephemeris in \S 3), the radial velocities of 
the two components with the corresponding standard deviations.}
\begin{tabular}{@{}ccccccccc@{}c}
\hline
HJD 2400000+ & Phase & \multicolumn{2}{c}{Star 1 }& \multicolumn{2}{c}{Star 2 } 	\\
             &       & $V_p$                      & $\sigma$                    & $V_s$   	& $\sigma$	\\
\hline
54908.5442	&0.8760		&	 51	&	1	&	-145	&	2		\\
54909.5455	&0.0800		&	-21	&	1	&	 133	&	2		\\
55346.5162	&0.0979		&	-21	&	3	&	 167    &	4	    	\\
55751.5070	&0.6010		&	 44	&	2	&	-131    &	7	  	\\
55796.4306	&0.7526		&	 72	&	2	&	-225	&	6	  	\\
55835.4714	&0.7059		&	 66	&	2	&	-220	&	4	  	\\
56130.5596	&0.8200		&	 77	&	2	&	-189	&	6	  	\\
56132.5791	&0.2314		&	-44	&	2	&	 248 	&	4	    	\\
56133.4902	&0.4171		&	-11	&	3	&	 123	&	7	    	\\
56134.6009	&0.6433		&	 56	&	3	&	-188	&	7	  	\\
56137.5467	&0.2434		&	-51	&	2	&	 255 	&	2	    	\\
56167.5138	&0.3482		&	-33	&	2	&	 188 	&	2	    	\\
56506.4260	&0.3900		&	-26	&	3	&	 154 	&	5	    	\\
56506.4516	&0.3952		&	-19	&	2	&	 134    &	6	    	\\
56507.4997	&0.6087		&	 55	&	2	&	-158	&	7	  	\\
56507.5210	&0.6131		&	 48	&	3	&	-156	&	8	  	\\
\hline \\
\end{tabular}
\end{minipage}
\end{table}

\begin{table}
\scriptsize
\centering
\begin{minipage}{85mm}
\caption {Orbital solution of the Algol--type binary in GT\,Cep.}
\begin{tabular}{@{}ccccccccc@{}c}
\hline
Parameter  & \multicolumn{2}{c}{\sf GT\,Cep }&  	\\
	   & Primary           & Secondary   		\\
\hline
 K (km s$^{-1}$) 			& 57 $\pm$ 2 	     	    	&236 $\pm$ 4		\\
 V$_\gamma$(km s$^{-1}$)          	& \multicolumn{2}{c}{$11 \pm 1$}            		\\
 Average O-C (km s$^{-1}$)		& 1.6         	     		&2.5      		\\    
 M $sin^3 i$ (M$_{\odot}$)		& 10.30 $\pm$ 0.48	  	& 2.49$ \pm $0.14	\\  
 Mass ratio, $q$                       	& \multicolumn{2}{c}{$0.2415 \pm 0.0094$} 		\\
$a \sin i$ (\Rsun)                      & \multicolumn{2}{c}{$28.42\pm 0.43$}  			\\
\hline
\end{tabular}
\end{minipage}
\end{table}

\subsection{Determination of the atmospheric parameters}
Mid-resolution optical spectroscopy permits us to derive most of the fundamental stellar parameters, such as  
projected rotational velocity ($V\,sin\,i$), spectral type (S$_p$), luminosity class, effective temperature 
(T$_{\rm eff}$), surface gravity ($log~g$), and metallicity ([Fe/H]).

The width of the cross-correlation function (CCF, hereafter) is a good tool for the measurement of projected 
rotational velocity ($V\sin i$) of a star. I use a method developed by \citet{Pen96} to estimate the $V\,\sin\,i$
of each star composing the investigated SB2 system from its CCF peak by a proper calibration based on a spectrum
of a narrow-lined star with similar spectral type. 
For the system, the rotational velocities of the components were obtained by measuring the FWHM of the CCF peak 
related to each component in five high-S/N spectra acquired near the quadratures, where the spectral lines have the 
largest Doppler-shift. The CCFs were used for the determination of $V\,sin\,i$ through a calibration of the full-width at 
half maximum (FWHM) of the CCF peak as a function of the $v\,sin\,i$ of artificially broadened spectra of slowly
rotating standard star (21\,Peg, $V\,sin\,i$=14\,\kms, e.g., \citet{Roy02}) acquired with the same setup and in the same observing
night as the target systems. The limb darkening coefficient was fixed at the theoretically predicted values, 0.42 for both 
components \citet{Van93}. We calibrated the relationship between the CCF Gaussian width and $V\sin i$ using the \citet{Con77} 
data sample. This analysis yielded projected rotational velocities for the components of GT\,Cep as $V_p\,sin\,i$=70\,\kms, and 
$V_s\,sin\,i$=70\,\kms.  The mean deviations were 4 and 7 \kms, for the primary and secondary, respectively, between the
measured velocities for different lines.

We also performed a spectral classification for the components of the system using COMPO2, an IDL code for the analysis of
high-resolution spectra of SB2 systems (see, e.g., Cakirli et al. 2014) and adapted to the TFOSC spectra. This
code searches for the best combination of two reference spectra able to reproduce the observed spectrum of the system. We give, as
input parameters, the radial velocities and projected rotational velocities $v\,sin\,i$ of the two components, which were
already derived. The code then finds, for the selected spectral region, the spectral types and fractional flux contributions that better
reproduce the observed spectrum, i.e. which minimize the residuals in the collection of difference (observed−composite) spectra.
For this task we used reference spectra taken from the \citet{Val04} Indo − U.S. Library of Coude Feed Stellar Spectra
(with a a resolution of $\approx$\,1\AA) that are representative of stars with various metallicity type, spectral types from late-O type 
to early-A, and luminosity classes V, IV, and III. The atmospheric parameters of these reference stars were recently revised by \citet{wu}. We
selected 198 reference spectra spanning the ranges of expected atmospheric parameters, which means that we have searched
for the best combination of spectra among 39204 possibilities per each spectrum. The observed spectra of GT\,Cep in the 
$\lambda\lambda$ 6525 –-- 6720 spectral region were best represented by the combination of HD\,886 (B2\,V) and HD\,77350 (A0\,IV). We 
have derived a spectral type of B2 main sequence star for the primary and A0 sub-giant for the secondary star of GT\,Cep, with an 
uncertainty of about 0.5 spectral subclass, by adopting the spectral type and luminosity class which are more frequently encountered. The 
effective temperature and surface gravity of the two components of each system are obtained as the weighted average of the values of the
best combinations of templates adopting a weight $w_i$ = 1/$\sigma_i^2$, where $\sigma_i$ is the average of residuals for the $i$-th combination.
The standard error of the weighted mean was adopted for the atmospheric parameters. Both stars appear to have a solar metallicity, within the 
errors. The atmospheric parameters obtained by the code and their standard errors are reported in Table\,5. The observed spectra of GT\,Cep 
at phases near to the quadratures are shown in Fig.\,3 \& 4 together with the combination of two reference spectra which give the best match.
The arrows in the top right panel of Fig.4 show C\,{\sc ii} $\lambda\lambda$ 6578 and 6583 lines.

\begin{table}
\scriptsize
\centering
\begin{minipage}{85mm}
\caption {Spectral types, effective temperatures, surface gravities, and rotational velocities of each components 
estimated from the spectra of GT\,Cep.}
\begin{tabular}{@{}lcccccccc@{}c}
\hline
Parameter  & \multicolumn{2}{c}{GT\,Cep }&  			\\
	   & Primary          	         & Secondary   		\\
\hline
 Spectral type 			& B2\,V 	     	&A0\,IV			\\
 T$_{\rm eff}$ (K)	    	& 22\,400$\pm$950  	&10\,100$\pm$650	\\   
 $\log~g$ ($cgs$)		& 3.86$\pm$0.05        	&3.16$\pm$0.17  	\\    
 $Vsin~i$ (km s$^{-1}$)  	& 70$\pm$4	  	&70$\pm$7     		\\ 
\hline
\end{tabular}
\end{minipage}
\end{table}

\begin{figure}
  \begin{center}
   \includegraphics[width=12.5cm]{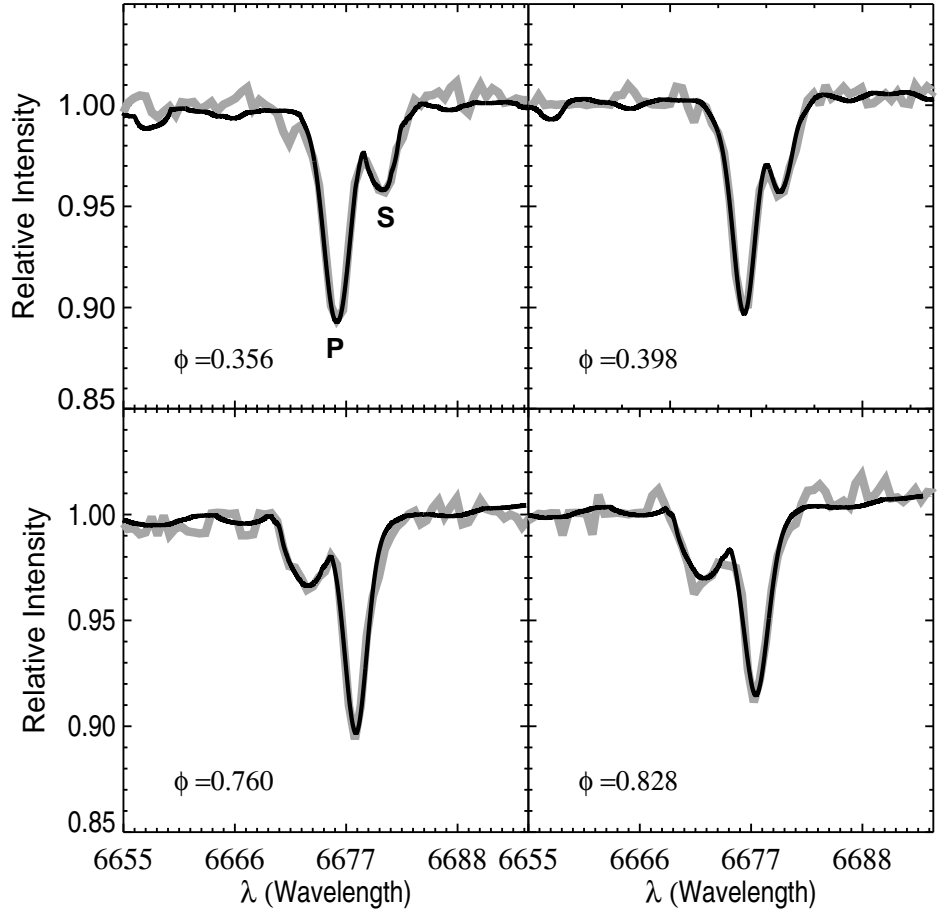}
   \end{center}
  \caption{Four spectra of GT\,Cep near opposing quadrature phases. The wavelength limits are 
  6650-6700 \AA, which include the He\,{\sc i} $\lambda$6678 line. The deeper lines in each 
  spectra refer to the primary star (P) and the shallower lines to the secondary (S). Vertical axis is the normalized flux.} 
\end{figure}

\begin{figure}
  \begin{center}
  \includegraphics[width=12.5cm]{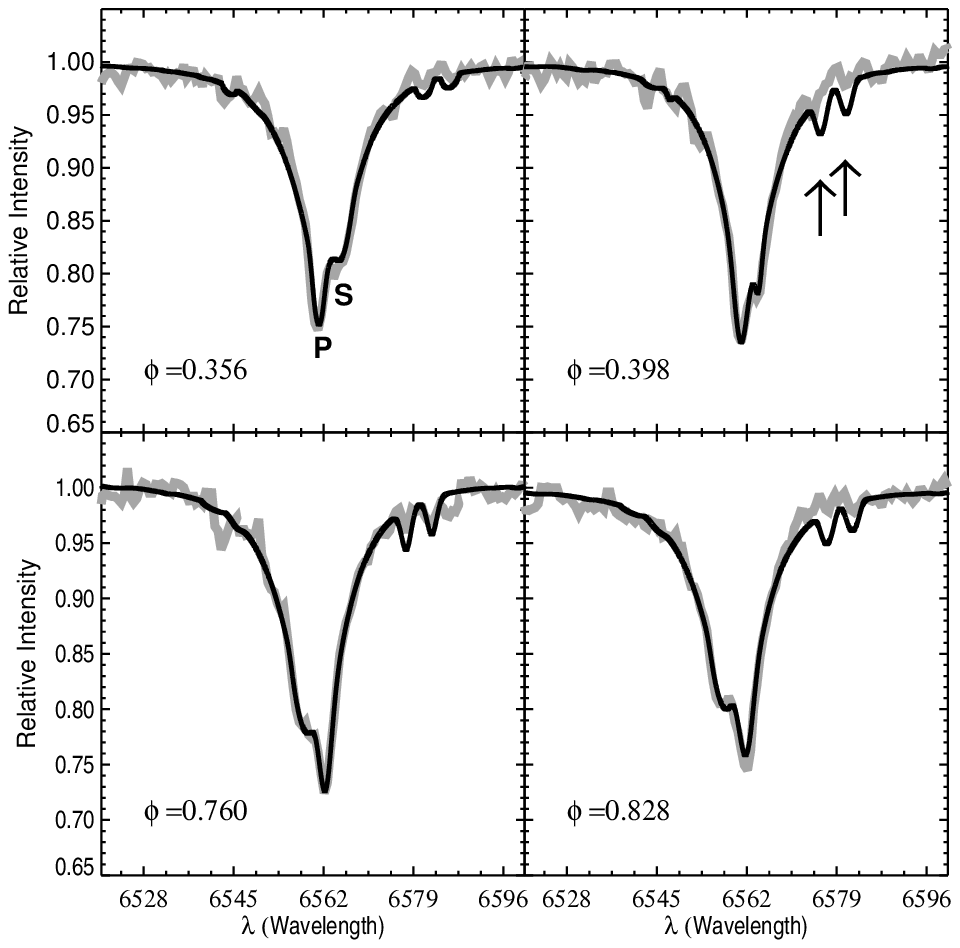}
   \end{center}
  \caption{Four spectra of GT\,Cep near opposing quadrature phases. The wavelength limits are 
  6500-6600 \AA, which include the H$\alpha$ line. The deeper lines in each spectra refer to 
  the primary star (P) and the shallowers to the secondary star (S). Vertical axis is the normalized flux.} 
\end{figure}

\section{ANALYSES OF THE LIGHT CURVES}
The photographic BV observations were collected by \citet{1975PZP.....2..171B}. The first photometric elements of the system were
calculated using the express method of Tsesesvich. Later on, the system was observed photoelectrically by \citet{1984A&AS...55..403B}.
The almost complete UBV light curves were obtained and analyzed by the methods $Russell-Merrill$, $Kitamura$ and $Wilson-Devinney$.
Recent light curves of the system were collected in the framework of two important surveys: Hp--band data of \citep{Per97} and R-band 
data of the NSVS project \citep{Wozniak04}. These magnitudes were obtained in a time interval 
of about three years. The accuracy of the Hipparcos data is about $\sigma_{H_p}$ $\sim$ 0.01\,mag. The $H_p$ magnitudes measured
by the $Hipparcos$ mission were transformed to the Johnson's V-passband using the transformation coefficients given
by \citet{Har98}.  All the data obtained by several researchers or surveys are plotted against the orbital phase in Fig.\,5. 

\begin{table*}
\scriptsize
\caption{Results of the analyses of the light curves for GT\,Cep.}
\begin{tabular}{lcccc}
\hline
Parameter	&Bartolini et al (1984)   &Bondarenko and Tokareva (1975) &Hipparcos & NSVS	\\
		& U\,B\,V 		  &	B\,V			  & V        & R	\\
\hline	
$i^{o}$			& 80.96$\pm$0.14     	&  81.142$\pm$0.21  		&86.91$\pm$0.24   	&85.72$\pm$0.50		\\
T$_{\rm eff_1}$ (K)	& 22\,400[Fix]		&  22\,400[Fix]	      		&22\,400[Fix]     	&22\,400[Fix] 		\\
T$_{\rm eff_2}$ (K)	& 10\,900$\pm$160	&  11\,000$\pm$200 		&10\,540$\pm$260  	&11\,520$\pm$370	\\
$\Omega_1$		& 4.476$\pm$0.090  	&  3.359$\pm$0.049  		&3.456$\pm$0.024  	&3.994$\pm$0.150	\\
$\Omega_2$		& 2.332$\pm$0.132     	&  2.332$\pm$0.132    		&2.332$\pm$0.132  	&2.333$\pm$0.112	\\
$r_1$			& 0.2374$\pm$0.0057  	&  0.3191$\pm$0.0054  		&0.3151$\pm$0.0025	&0.2678$\pm$0.0109	\\
$r_2$			& 0.2626$\pm$0.0065  	&  0.2625$\pm$0.0044		&0.2625$\pm$0.0021	&0.2625$\pm$0.0107	\\
$\frac{L_{1}}{(L_{1}+L_{2})}$R 		&    ---           	&       ---           	&    ---        &0.7491$\pm$0.0212	\\
$\frac{L_{1}}{(L_{1}+L_{2})}$V 		& 0.7655$\pm$0.0109	&  0.8766$\pm$0.0082	&0.847$\pm$0.006&0.7491$\pm$0.0212	\\
$\frac{L_{1}}{(L_{1}+L_{2})}$B 		& 0.7505$\pm$0.0128	&  0.8006$\pm$0.0084	& ---           &---	\\
$\frac{L_{1}}{(L_{1}+L_{2})}$U		& 0.8302$\pm$0.0124	&  ---                 	& ---           &---	\\
$\sum(O-C)^{2}$				& 0.848             	&  0.987	       	&0.987	        &0.988	\\
$N$					& 1446             	&  532		     	&134	        &107	\\
$\sigma$				& 0.008             	&  0.091     		&0.087	        &0.098	\\
\hline
\end{tabular}
\end{table*}
\begin{figure}
\center
\includegraphics[width=8cm,angle=0]{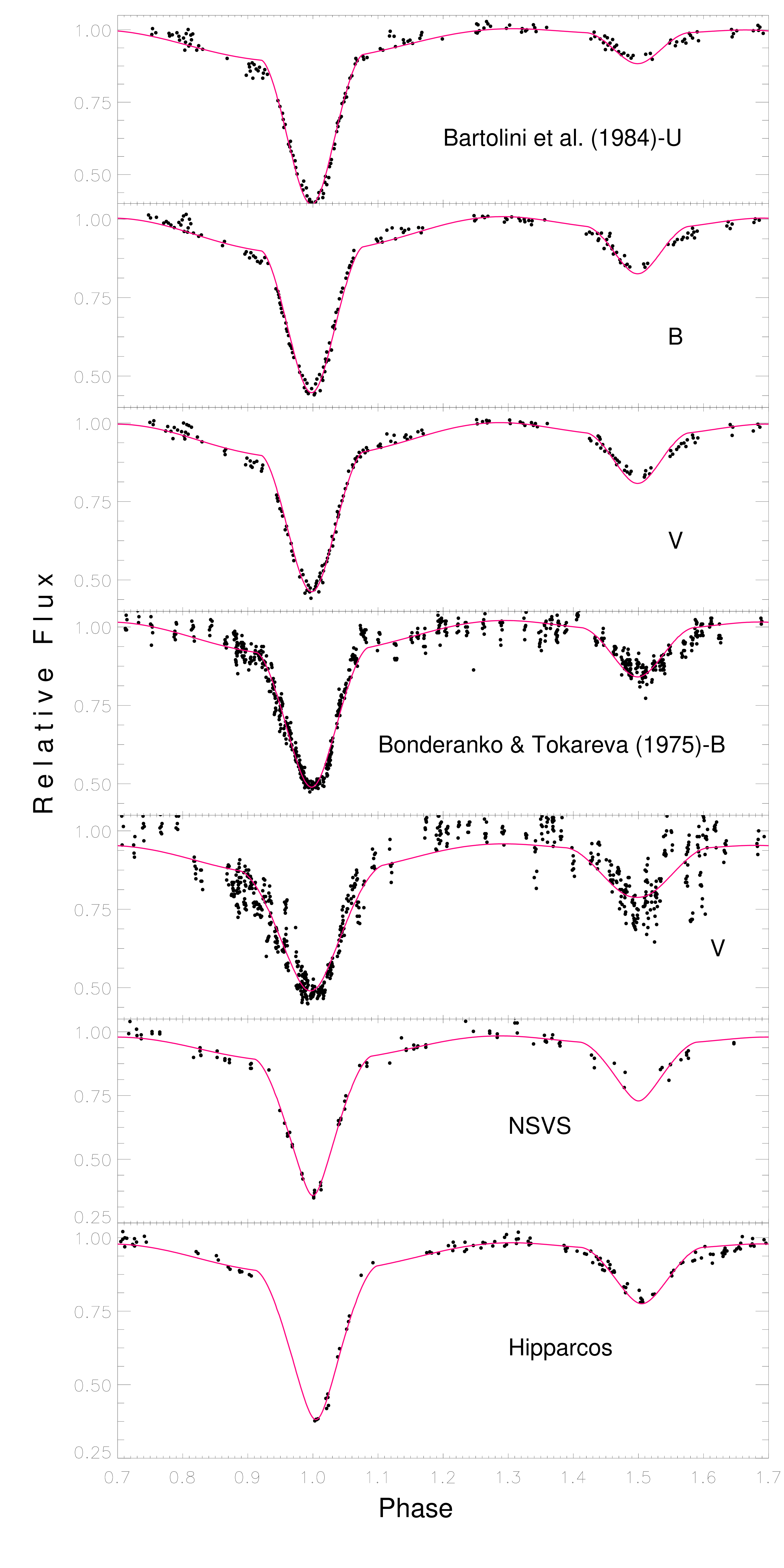}
\caption{Comparison of the observed and computed light curves of GT\,Cep. The continuous lines show the best-fit model.} \end{figure}

The most--commonly used code for modelling the light curves of the eclipsing binaries is that of \citet{Wilson71}. This 
code was up-dated and implemented in the {\sc phoebe} code of \citet{Prs05}. One of the main difficulties
in this modelling is determination of individual effective temperatures of both stars. Generally used practice is 
to estimate effective temperature of primary star and determine that of the secondary star. Effective temperature of 
the primary star could be estimated from its spectra or color indices. The average apparent visual magnitude
and colors are measured by \citet{Lac92} as V=8.125$\pm$0.017, U-B=-0.532$\pm$0.013, B-V=0.339$\pm$0.003.
The quantity $Q$=($U-B$)-($E_{(U-B)}$/($E_{(B-V)}$)$(B-V)$ is independent of interstellar 
extinction. The average value obtained is ($E_{(U-B)}$/($E_{(B-V)}$)=0.72$\pm$0.03 \citet{Joh53}; \citet{Hov04}. We compute the 
reddening-free index as $Q$=-0.776$\pm$0.013. The values of index were calculated by \citet{Hov04} begining from O8 to G2 
spectral type for the luminosity classes between main-sequence and supergiants. A calibration between $Q$ and spectral 
type yield a spectral type of B1\,V, in agreement with spectral classification made from the spectra. In addition the 
infrared colors J-H=0.096$\pm$0.05, H-K=-0.007$\pm$0.05\,mag are given in the {\sc 2MASS} catalogue \citep{Cut03} 
correspond to an early B type star.

Logarithmic limb-darkening coefficients were interpolated from the tables of \citet{Van93}. They are updated at every 
iteration. The gravity-brightening coefficients $g_1$=$g_2$=1.0 and albedos $A_1$=$A_2$=1.0 were fixed for both 
components, as appropriate for stars with radiative atmospheres. Since the preliminary analysis indicates that secondary 
star fills its Roche lobe, synchronous rotations were adopted and Mode\,5 was used in the solution. This mode is used 
for the Algol systems, e.g., secondary star fill their limiting Roche lobes. The BV light 
curves of \citet{1975PZP.....2..171B}, UBV light curves of \citet{1984A&AS...55..403B}, $Hipparcos$ and $NSVS$ 
were analyzed separately.

The adjustable parameters in the light curves fitting were the orbital inclination $i$, the effective temperature 
of the secondary star T$_{\rm eff_2}$, the luminosity of the primary L$_1$, and the zero-epoch offset. The parameters 
of our final solution are listed in Table\,6. The uncertainties assigned to the adjusted parameters are the internal 
errors provided directly by the code. In the last three lines of Table\,6 sums of squares of residuals $\sum(O-C)^{2}$, 
number of data points $N$, and standard deviations $\sigma$ of the observed light curves are presented, respectively. It 
is obvious that the $\sigma$-value obtained for the UBV light curves of \citet{1984A&AS...55..403B} is the smallest, 
amounting to at least eleventh of the others. Therefore I adopt the parameters obtained from the UBV light curves as 
the best fit model parameters for GT\,Cep. The O-C residuals point out a sinusoidal change which has been
attributed to a third-body orbit. We repeated the analysis taking the third-light as an adjustable parameter.
The analysis showed that there is no sign about the existence of the light contribution of an additional component.
The computed light curves are compared with the observations in Fig.\,5.

\begin{table}
\scriptsize
\centering
%\begin{minipage}{85mm}
\caption{Absolute parameters, magnitudes and colours for the components of GT\,Cep. }
\begin{tabular}{@{}lcccccccc@{}c}
\hline
Parameter  		& \multicolumn{2}{c}{\sf GT\,Cep }&		   	\\
			& Primary              		  & Secondary		\\
\hline
Mass (M$_{\odot}$)	& 10.70$\pm$0.50	 	& 2.58$\pm$0.14       	\\ 
Radius (R$_{\odot}$)	& 6.34$\pm$0.19	 		& 6.98$\pm$0.11 	\\
$T_{\rm eff}$ (K)	& 22\,400$\pm$950       	& 10\,900$\pm$300	\\
$\log~(L/L_{\odot})$	& 4.025$\pm$0.078       	& 2.862$\pm$0.035	\\      
$\log~g$ ($cgs$)	& 3.798$\pm$0.022       	& 3.094$\pm$0.025	\\
$Sp.Type$		& B2\,V		 		& A0\,IV		\\
$M_{bol}$ (mag) 	& -5.31$\pm$0.19		& -2.40$\pm$0.09 	\\
$BC$ (mag)		& -2.18		 		& -0.43	     	     	\\
$M_{V}$ (mag)		& -3.14$\pm$0.10	 	&-1.97$\pm$0.07 	\\
$(Vsin~i)_{calc.}$ (km s$^{-1}$)	& 70$\pm$2	& 78$\pm$2     	     	\\
$(Vsin~i)_{obs.}$ (km s$^{-1}$) 	& 70$\pm$4	& 70$\pm$7     	     	\\
$d$ (pc)				& 854$\pm$43	& 903$\pm$52 	        \\
\hline
\end{tabular}
\medskip
%\end{minipage}
\end{table}

\section{RESULTS AND DISCUSSION}
Based on mid$-$resolution spectroscopic observations I have obtained radial velocities of both components for the high$-$mass 
eclipsing binary GT\,Cep. Analysis of the radial velocities yielded M$_1$ sin$^3$$i$=10.30 M$_{\odot}$, M$_2$ sin$^3$$i$=2.49 
M$_{\odot}$, and $a \sin i$=28.42 \Rsun. These values are too different from those of the previous determination by 
\citet{pim1964}. Combining the results of the simultaneous UBV light curves' analysis, listed in the second column of 
Table\,6, I determined absolute parameters for the components. For calculation of the fundamental stellar parameters for 
the components such as masses, radii, luminosities and their formal standard deviations the JKTABSDIM\footnote{This can 
be obtained from http://http://www.astro.keele.ac. uk/$\sim$jkt/codes.html} code was used. This code calculates physical 
parameters and distance to the system using several different sources of bolometric corrections \citep{Sou05}. The final 
results are presented in Table\,7.

The luminosities of the components were calculated from their absolute radii and effective temperatures. Using these 
luminosities I have calculated bolometric absolute magnitudes of the components, assuming solar bolometric absolute 
magnitude of 4.74\,mag. Next I applied a bolometric correction for each star based on our spectral type and the effective 
temperatures, listed in Table\,7, from \citep{Flo96} to obtain their absolute visual magnitudes. Analysis of the light 
curve yields light ratio of 0.306 for the V-passband. This light ratio, the observed colours and the intrinsic colour 
of the primary star of $(B-V)_0$=$-$0.25$\pm$0.01 \citep{Dri00} allow us to estimate an intrinsic composite colour of 
$(B-V)_0$=$-$0.27$\pm$0.01. Thus, the interstellar reddening of $E_{(B-V)}$=0.61$\pm$0.01\,mag  and absorption in the 
V-passband of $A_{V)}$=1.89$\pm$0.01\,mag  are estimated for the system. I arrived at distance modulus for the primary 
and secondary component as 9.66 and 9.78\,mag, respectively, by using $V_{1}$=8.42 and $V_{2}$=9.70\,mag. These distance 
moduli correspond to a distance of 854$\pm$43 pc for the primary and 903$\pm$52 pc for the secondary star. Since the 
secondary star is an evolved star and losing its mass I adopt distance to the system estimated from the primary component.

\begin{figure}  
\includegraphics[scale=1.05]{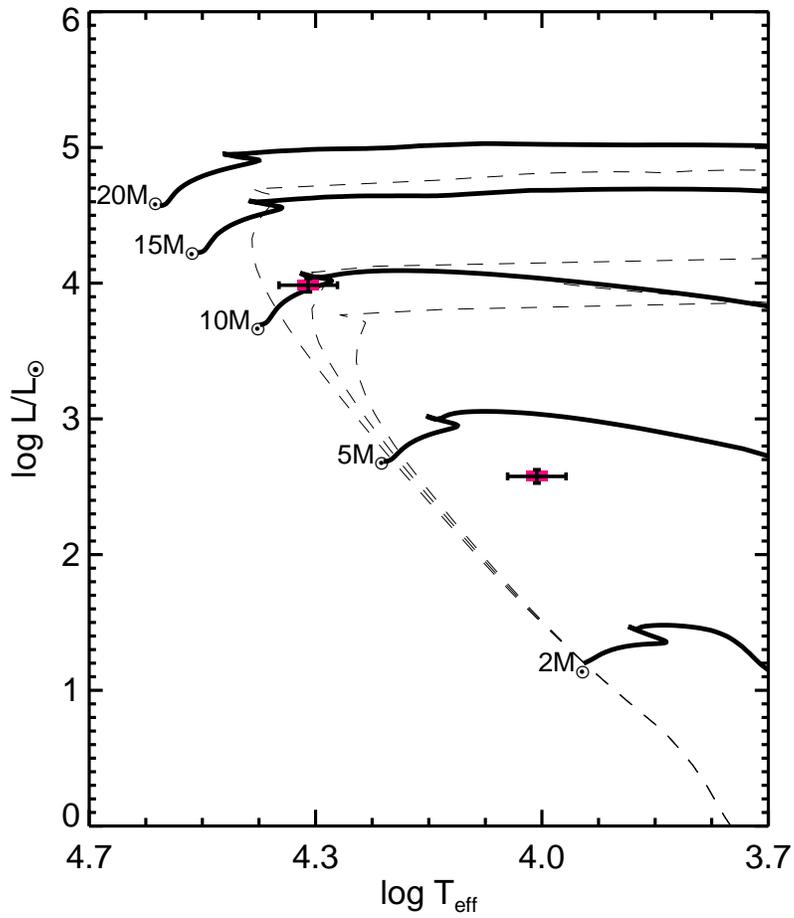}
\caption{Locations of the components on the effective temperature-luminosity panel. Evolutionary tracks (solid lines) for 
stars 2, 5, 10, 15 and 20\,M$_{\odot}$ and isochrones  correspond to 10, 20, and 30\,Myr (dashed lines, going from left to 
right) for single rotating stars with solar metallicity \citep{Ekstrom12}. } \end{figure}

Figure\,6 shows the components, with 1-$sigma$ error bars, of GT\,Cep in the log T$_{\rm eff}$-log L/L$_{\odot}$ panel. 
We used the models of \citet{Ekstrom12}, for single stars in the range 0.8$-$120\,M$_{\odot}$ at solar metallicity 
(Z=0.014), which include rotation to comparison of our measured parameters with evolutionary models. While these 
evolutionary models are not appropriate for the SDABs, I plot the stars in the well-known Hertzsprung-Russell diagram 
to show their locations.

In the short-period, P$<5$ d, SDABs the mass-gaining components are large enough, relative to the separation, that the 
infalling mass can directly impact on the accretor. For the first time, \citet{Lub75} discussed and modelled formation 
of discs in semi-detached systems. In Fig.3 of \citet{Der10}, the so-called $r-q$ diagram, the SDABs  with and without 
discs are plotted. With a fractional radius of 0.24 and a mass-ratio of 0.24, GT\,Cep locates in the $r-q$ diagram where 
the gainers of SDABs without permanent or transient disc are gathered. In Fig.2 of \citet{Der10} the ratios of the observed equatorial 
to the computed synchronous rotational velocities for the SDABs were plotted against the mass of the gainers. The 
mass-gaining primary star of GT\,Cep is appeared as the highest mass star with synchronous rotational velocities. \citet{Ren06} studied 
evolution of interacting binaries with a B type primary at birth. They have taken into account both loss of mass and angular momentum 
during binary evolution. Their calculation clearly showed that the binaries have to lose a significant amount of mass without losing 
much angular momentum. Their liberal scenario with much mass loss without much loss of angular momentum yielded orbital periods and
mass ratios of Algols that are in better agreement with the observations. Though physical events related to the liberal scenario 
at binaries with a B type binary at birth are yet not fully understood they have estimated that only half of the matter lost by 
the donor is captured by the gainer and the other half left the system. This assumption reveals that a mass of about 3\,M$_{\odot}$ 
is lost for the case of GT\,Cep during its evolution.

Since we found the period and projected radius of the orbit the mass of the third star can only be computed for 
different values of orbital inclination. The total mass of the eclipsing pair was obtained as 13.28 M$_{\odot}$.
For the inclination of the third-body orbit of $90^{o}$, $80^{o}$ and $70^{o}$ I calculate masses for the
third star as 6.9, 7.0 and 7.45 M$_{\odot}$, respectively. For a wide range of orbital inclination the mass
of third star is about 7 solar masses. If it were a main-sequence star its visual magnitude would be 8.8 mag, i.e. 0.4\,mag 
fainter than the primary and 0.9\,mag brighter than the secondary component. The third star would be about 0.7\,mag fainter
from the eclipsing pair. Assuming the distance to the eclipsing pair as 854\,pc, and separation between the eclipsing 
pair and the third star as 40\,$AU$ one can easily calculate an angular separation of about 0.05 arcseconds. Such a star could be 
angularly resolved by contemporary instruments and methods.  Since no photometric or spectroscopic
signature has been observed for the third star up to date it would be reasonable to think that this component either 
might not be a normal star or it is a multiple star system.

%%%%%%%%%%%%%%%%%%%%%%%%%%%%%%%%%%%%%%%%%%%%%%%%%%%%%%%%%%%%%%%%%%%%%%%%%%%%%%%%%%%%%%
\section*{Acknowledgments}
This paper is dedicated to Professor C. Ibanoglu who spent most his life on the structure and evolution of the binary 
stars. He surveyed, for example, more than thirty years the exotic binary V471\,Tauri. He and his colleagues carried 
out many quantitative studies of the physical properties of the close binaries. I am grateful for his aid in every 
step of my studies and his valuable contributions to close binary systems.
We thank to T\"{U}B{\.I}TAK National Observatory (TUG) for a partial support in using RTT150 
telescope with project number 11BRTT150-198.
This study is supported by Turkish Scientific and Technology Council under project number 112T263.
We thank to EB{\.I}LTEM Ege University Research Center for a partial support with project number {\sf 2013/BIL/018}.
The following internet-based resources were used in research for this paper: the NASA Astrophysics Data 
System; the SIMBAD database operated at CDS, Strasbourg, France; and the ar$\chi$iv scientific paper 
preprint service operated by Cornell University. 
%%%%%%%%%%%%%%%%%%%%%%%%%%%%%%%%%%%%%%%%%%%%%%%%%%%%%%%%%%%%%%%%%%%%%%%%%%%%%%

\end{document}